\documentstyle[aps,12pt,preprint]{revtex}

\begin{document}
\title{History effect in inhomogeneous superconductors}
\author{Y. Liu, H. Luo, X. Leng, Z. H. Wang, L. Qiu, S. Y. Ding$^{*}$}
\address{Department of Physics and National Laboratory of Solid State\\
Microstructures, Nanjing University, Nanjing 210093, P. R. China}
\author{L. Z. Lin}
\address{Institute of Electric Engineering, Chinese Academy of Science, Beijing\\
100080, P. R. China}
\maketitle

\begin{abstract}
A model was proposed to account for a new kind of history effect in the
transport measurement of a sample with inhomogeneous flux pinning coupled
with flux creep. The inhomogeneity of flux pinning was described in terms of
alternating weak pinning (lower j$_c$) and strong pinning region (higher j$%
_c $). The flux creep was characterized by logarithmic barrier. Based on
this model, we numerically observed the same clockwise V-I loops as reported
in references. Moreover, we predicted behaviors of the V-I loop at different
sweeping rates of applied current dI/dt and magnetic fields B$_a$, etc.
Electric transport measurement was performed in Ag-sheathed Bi$_{2-x}$Pb$_x$%
Sr$_2$Ca$_2$Cu$_3$O$_y$ tapes immersed in liquid nitrogen with and without
magnetic fields. V-I loop at certain dI/dt and B$_a$ was observed. It is
found that the area of the loop is more sensitive to dI/dt than to B$_a$,
which is in agreement well with our numerical results.
\end{abstract}



$*$ Corresponding author. {\bf Email: }syding@netra.nju.edu.cn

\today
\newpage

\section{Introduction}

\vspace{0in} Recently the so-called history effect (HE) of vortex matter has
been frequently observed by a variety of methods such as electric transport
measurement, dc magnetization hysteresis, ac susceptibility and torque
techniques, and drawn much more attention \cite{1,2,3,4,5,6,7,8,9,10}. HE
refers to that critical current density j$_c$ of a superconductor has
different values at a same field (or temperature) for either a thermal,
magnetic or current cycles. The earlier observed hysteresis of R-T curves in
the thermal cycle in fact is a kind of HE caused by first order phase
transition of vortex matter (melting or freezing) \cite{11}. Different
mechanisms have been proposed to account for HE \cite{1,2,3,4,5,6,7,8,9,10}.
And a typical explanation contains two assumptions: relatively weak flux
pinning and uniform small quenched disorders. Then HE is explained with a
theory of disorder-order transition of vortex matter. The disordered vortex
matter pinned more strongly (higher j$_c$) is supercooled to low temperature
where it is in metastable state and then a process such as a bias current
anneals the metastable disordered state into an ordered and stable one with
lower j$_c$. An important point is that flux creep in the models is omitted
for the low temperatures. In this case, the property of vortex matter is
governed only by the competition between interactions of vortices-vortices
and vortices-quenched disorders.

However, a different kind of HE has been observed \cite{12}. The hysteresis
loop of V-I curve of polycrystalline Bi$_{2-x}$Pb$_x$Sr$_2$Ca$_2$Cu$_3$O$_y$
(Bi2223) at liquid nitrogen temperature was measured there. See also Fig.6
in this paper. It is clear that such a V-I loop is a new kind of HE and very
different from the V-I loop in low temperature superconductors reported
before \cite{1}. The difference between these two kinds of HE is that the
direction of the V-I loop is anti-clockwise in reference 1 while clockwise
in reference 12 and this paper. Despite the anti-clockwise loop has been
explained very well, the clockwise one has not yet been understood up to our
knowledge and interested us very much.

Note that this new kind of HE takes place in a system different from those
with uniform and weak quenched disorders at low temperatures. First, at
higher temperatures, such as liquid nitrogen temperatures, thermal
fluctuation should be important and flux creep could influence the property
of vortex matter apparently \cite{13}. In fact the influence of flux creep
on the V-I curve and transport property of Bi2223 have been confirmed by
earlier experimental and numerical results \cite{14,15,16,17,18}. Second,
the quenched disorders or flux pinnings are inhomogeneous in polycrystalline
high temperature superconductors (HTS). There are weak links and grains in
sintered samples such as polycrystalline Bi2223. And the flux pinning
strength of the weak links at grain boundaries is very different from that
of the grains \cite{19}. In view of these two points, any models concerning
the system with homogenous quenched disorders without flux creep could not
be used to understand this novel hysteresis behavior in V-I curve.

A theoretical model describing irreversible electric-magnetic behaviors of a
system with inhomogeneous quenched disorders has already been proposed \cite
{20}. Unfortunately, flux creep was omitted there. Moreover, this model has
not been used to explain any HE, including the clockwise V-I loop.
Experimentally, A.M. Campbell's group did many important works on
inhomogeneous superconductors. They did observe the hysteresis V-I curves
before. However, only a qualitative explanation was given. And what
interests us now is how to simulated the clockwise V-I loop.

Considering flux creep and inhomogeneous flux pinning, we proposed a new
model to explain this new kind of HE. We carried out numerical simulation
with this model. Upon explaining the clockwise loop of V-I curve reported in
reference 12, we further predicted some new phenomena. To test the
predictions, electric transport measurement was conducted for characteristic
V-I curve of silver-sheathed Bi$_{2-x}$Pb$_x$Sr$_2$Ca$_2$Cu$_3$O$_y$ tapes
(Ag-Bi2223) immersed in liquid nitrogen with or without applied fields B$_a$%
. Clockwise loop of V-I curve at certain dI/dt and B$_a$ was observed and is
dependent on dI/dt and B$_a$, confirming the numerical result.

\section{Model}

\smallskip The model consists of two basic assumptions: there exists flux
creep in a sample and the flux pinning is inhomogeneous. The complex
inhomogeneity of flux pinning is simplified by periodic strong pinning
region (S) with large j$_c$ (j$_{cS}$) and weak pinning region (W) with
small j$_c$ (j$_{cW}$), see Fig.1. The flux diffusion is characterized by
collective creep of vortex glass, which has been discussed widely \cite{21}.
To simplify the calculation, we use logarithmic U(j) relationship, which is
the special case of vortex glass model with very small glass exponent $\mu $%
\cite{22}.

We should emphasize that our model is not a simplified version of the
''brickwall'' model, which has been widely discussed before \cite{23}. There
are some resemblances between them. However, there are at least two points
that distinguish our model from the previous one. First, the ''brickwall''
model is a static critical state model without flux creep. Obviously, in any
static models, the transport properties studied here will be independent of
the sweeping rate of applied current dI/dt, contrasting with the
experimental observation. Second, in the ''brickwall'' model the S and W
regions are connected in series along the direction of current. While in our
model, they are connected in parallel.

In our one-dimensional model, flux lines firstly enter the system from the
surfaces into the weak pinning channel (W) and then diffuse into the strong
pinning region (S). So W surrounds S. In other words, S is inside the
slablike sample.

As an example, the model may be proper to polycrystalline samples of HTS,
such as silver-sheathed Bi$_{2-x}$Pb$_x$Sr$_2$Ca$_2$Cu$_3$O$_y$ (Ag-Bi2223).
The grain boundaries in polycrystalline HTS form weak link network of flux
pinning and thus can be considered as W, whereas the intra-grain region is
S. It is well known that the weak link network suppresses substantially the
critical current density of HTS such as Ag-Bi2223. When current or magnetic
field is applied, flux density will enter inter-grain (W) at first and then
penetrate gradually into the intra-grain (S) by means of flux creep. Another
example the model may be applicable to is the low temperature
superconductors, such as Nb$_3$Sn metallic compounds, where flux lines may
be pinned by grain boundary (S) whereas the intra-grain is weaker pinning
region (W) \cite{24}.

Certainly, one cannot expect that numerical result by such a simplified
phenomenological model can quantitatively represent experimental data of a
sample. For example, the one-dimensional geometry assumption in the model is
not fulfilled for a real sample. And, the flux pinning is more complex by
far than the two model parameters, j$_{cS}$ and j$_{cW}$. Nevertheless, our
model is a new one accounting for the experimental clockwise V-I loop, which
was observed very recently. The effectiveness of this model was demonstrated
by the electric transport measurement of hysteresis V-I loop.

\section{Simulation}

\subsection{Basic Equation}

\smallskip Consider a slab with infinite length along y-axis, thickness d
along the x-axis and width w along the z-axis. The current is applied along
the y-axis. In view of w%
\mbox{$>$}%
\mbox{$>$}%
d, we focus on the one-dimensional case for simplicity. The non-linear
diffusion equation describing the macroscopic B or j is well known \cite
{25,26}. With a logarithmic dependence of U on j: $U(j)=U_0\ln \left|
j_c/j\right| $, the diffusion equation of flux can be written as:

\begin{equation}
\frac{\partial B}{\partial t}=\frac{v_0}{(\mu _0j_c)^{n+1}}\frac \partial {%
\partial x}\left[ \left| \frac{\partial B}{\partial x}\right| ^n\left( \frac{%
\partial B}{\partial x}\right) B\right]  \eqnum{1}
\end{equation}

where $n=U_0/kT$ and $v_0=u\omega _m$. $v_0$ is the attempt velocity of the
thermal-activated vortex motion, $u$ is the hopping distance and $\omega _m$
is the microscopic attempt frequency.

\subsection{Boundary and initial conditions}

In the V-I curve measurement, the current is always applied with a certain
sweeping rate dI/dt. The boundary condition of Eq.1 can be obtained from the
current conservation equation: $\frac \partial {\partial t}%
\int_0^w\int_0^djdxdz=\frac{dI}{dt}$. Substituting $\partial B/\partial
x=-\mu _0j$ into it, one comes $B(d,t)-B(0,t)=-\mu _0I/w.$ Due to
antisymmetry, one has $B(d,t)=-B(0,t)$, thus the boundary condition is 
\begin{equation}
B(0,t)=\frac{\mu _0I}{2w}=\left( \frac{\mu _0}{2w}\frac{dI}{dt}\right) t 
\eqnum{2}
\end{equation}
\newline
As for the initial condition, there is no current at t=0 in the sample, hence

\begin{equation}
B(x,0)=0  \eqnum{3}
\end{equation}

\subsection{Numerical method and the choice of parameters}

Such kind of non-linear diffusion equation can be numerically solved with
the finite difference method. First, the temporal and spatial variances are
discretized with different steps respectively. Since the temporal and
spatial steps are very important for the stability of the calculation, we
take pains to choose suitable steps to avoid a divergence problem. Second,
the equation is discretized and rewritten in an implicit difference scheme,
which is better than the explicit difference scheme in the stability of
calculation. Finally, based on the initial condition, the calculation by
iteration will obtain the temporal and spatial magnitude of physical
quantity, namely B(x,t), j(x,t) and E(x,t). The voltage can be obtained by
integration of E(x,t). Note that the boundary condition plays a role of
constraint in every time step during current ascending and descending, see
Eq.2.

It is easy to see that $U(j)=U_0\ln \left| j_c/j\right| $ combined with $%
E=E_0\exp (-U/kT)$ leads to the power dependence of the E-j curve $%
E=E_0(j/j_c)^n$ with $n=U_0/kT$. This power law E-j curves are observed
frequently in transport measurements and can be used to determine the
parameter $n$ \cite{15}. Certainly, $n$ is dependent on flux pinning
strength, magnetic field and temperature \cite{18,27}. The weaker flux
pinning and higher temperature bring about smaller n. In fact, many
experiments including the present work show that for the Ag-Bi2223 in liquid
nitrogen temperature, $n=6$ is a typical value.

As for the velocity of the thermal-activated vortex motion and the critical
current density, typical and reasonable magnitudes were employed based on
the earlier theoretical and experimental work, for example we took $%
v_0=u\omega _m=1m/s$ ( if $u$ $\symbol{126}10^{-6}m$, $\omega _m\symbol{126}%
10^6s^{-1}$) and $j_{cW}=2\times 10^8A/m^2$. It should be pointed out that
our numerical result is not sensitive to the choice of $n$, $v_0$ and j$%
_{cW} $.

\subsection{Numerical results and discussions}

\subsubsection{Effect of j$_{cS}$/j$_{cW}$ ratio on the hysteresis loop}

Since j$_c$ represents the pinning strength, the ratio j$_{cS}$/j$_{cW}$ can
be considered as a parameter reflecting the ratio of the two pinning
strengths. Shown in Fig.2 is our numerical result of the dependence of HE on
j$_{cS}$/j$_{cW}$ at fixed j$_{cW}$, d$_S$/d$_W$ and dI/dt. It is very clear
that the V-I loop is clockwise as reported in reference 12. To understand
how j$_{cS}$/j$_{cW}$ affects the V-I loop, the corresponding current
distributions in sample were also calculated.

Note that the velocity of flux diffusion, namely the velocity of current
diffusion in transport measurements, can be written as: $v=v_0\exp (-U/kT)$ $%
\propto (j/j_c)^n$ for logarithmic U(j). Hence, for the same $j$, the larger
the j$_c$, the smaller the $v$. Generally speaking, the average speed of
current diffusion in S region ( $\overline{v_s}$ ) is smaller than that of W
region ( $\overline{v_w}$ ). In our calculation, j$_{cW}$ is fixed, i.e. $%
\overline{v_w}$ is fixed. By changing j$_{cS}$/j$_{cW}$ ratio, namely j$%
_{cS} $, we adjust $\overline{v_s}$ only.

For a large ratio (j$_{cS}$/j$_{cW}$=10), i.e. high inhomogeneity of flux
pinning, $\overline{v_s}$ is so small comparing with $\overline{v_w}$. And
in such a case, there is no enough time for current diffusion in response to
both the ascending branch of V-I curve (Fig.2.1(b)) and the descending one
(Fig.2.1(c)). Consequently, no obvious V-I loop can be detected (Fig.2.1(a)).

For a medium ratio (j$_{cS}$/j$_{cW}$=4), i.e. a medium inhomogeneity of
flux pinning and a larger $\overline{v_s}$, it is shown that the area of the
loop is very large (Fig.2.2(a)). In this case, although current has already
penetrated into S region during the ascending branch (Fig.2.2 (b)), much
more current penetrates into S during the descending branch (Fig.2.2(c)).
That is to say, the current diffusion of S region has more time in response
to the descending branch than to the ascending one. As a result, the V-I
loop is obvious.

For a small j$_{cS}$/j$_{cW}$ ratio, low inhomogeneity of flux pinning, $%
\overline{v_s}$ is very large and approaches $\overline{v_w}$. And there is
little difference between the distributions of current during the two
branches (Fig.2.3(b), (c)). The minor difference is only seen when the
current is small and the corresponding voltage is too small to be detected.
Hence, there is almost no hysteresis loop (Fig.2.3 (a)). Especially, at the
uniform case (j$_{cS}$/j$_{cW}$ =1, $\overline{v_s}=\overline{v_w}$), there
will be no loop at all.

\subsubsection{Effect of d$_S$/d$_W$ ratio on the hysteresis loop}

With a fixed total thickness, the spatial distribution of S and W region,
i.e. the thickness ratio of them is also important for the V-I loop. The
numerical V-I loops with different d$_S$/d$_W$ at fixed dI/dt and j$_{cS}$/j$%
_{cW}$ were displayed in Fig.3. It is found that the smaller the d$_S$/d$_W$%
, the smaller the area of the loop.

For d$_S$/d$_W$ =0.857 and other parameters given here, the thickness of S
region i.e. the distance for current diffusion in S region is so long that
current diffusion in S region has less time to respond the ascending branch
(Fig.3.1(b)) than to respond the descending one (Fig.3.1(c)). So the area of
the V-I loop is large (Fig.3.1(a)).

For d$_S$/d$_W$ =0.222, S region is thinner, i.e. the distance of current
diffusion in S region is shorter, resulting that comparing with the d$_S$/d$%
_W$ =0.857 case, the current diffusion in S region has more time in response
to the ascending current (Fig.3.2(b)). In other words, for such parameters
given here, the response time of current diffusion in S region to the two
branches of the V-I loop shows little difference. As a result, the area of
the V-I loop becomes smaller (Fig.3.2(a)).

As for d$_S$/d$_W$ =0.105, S region is too thin and the distance of current
diffusion in S region is too short. Hence there is enough time for current
diffusion in response to both current ascending and descending at the given
parameters (Fig.3.2(b), (c)). That is to say, there is no obvious V-I loop
(Fig.3.3(a)). As d$_S$/d$_W$ approaches 0, namely the uniform flux pinning
case, the hysteresis loop will disappear thoroughly.

\subsubsection{Current sweeping rate (dI/dt) dependent V-I loop}

The numerical result of current sweeping rate dependent V-I loops at fixed j$%
_{cS}$/j$_{cW}$ and d$_S$/d$_W$ was illustrated in Fig.4.

It is seen that if dI/dt is very small, such as 0.01A/s, flux lines will
have enough time to respond the changing applied current and thus can
penetrate into both W and S regions. The difference of current distribution
during the ascending (Fig.4.1(b)) and descending branch (Fig.4.1(c)) is very
small. As a result, the area of the V-I loop is also small (Fig.4.1(a)). As
dI/dt reaches a moderate value, say dI/dt=2A/s, the difference between the
current distributions during the ascending and descending branch is larger.
And the area of the V-I loop becomes larger too(Fig.4.2(a)). Nevertheless,
when dI/dt is too large, say dI/dt=10A/s, the area of the V-I loop decreases
dramatically (Fig.4.3(a)). It is easy to understand that in such a case,
comparing with dI/dt, $\overline{v_s}$ is so small that the current cannot
penetrate into the S region during both the ascending and descending branch.
And the corresponding current distributions during the two branches do not
show visible difference (Fig.4.3(b), (c)). So there is no obvious V-I loop.

Now, it is clear what the effect of dI/dt on the V-I loop means. With fixed
distance and average speed of the current diffusion, to change dI/dt is
equivalent to shift our observing time window. When we choose time window by
using dI/dt as 10A/s, the observing time is so early that there is no enough
current inside S region by means of diffusion even the applied current
sweeping has been finished. Of course there is no detectable V-I loop. If we
use dI/dt=0.01A/s, we shift our time window so late that there are always
enough current diffused into S region during both the ascending and
descending branch. It is natural that no obvious V-I loop exists at all.
Only when we choose a proper observing window, i.e. a proper dI/dt such as
2A/s, is there much more current in S region during the descending branch
than the ascending one. In other words, only in such case, can the current
diffusion in S region has more time to respond the descending branch than
the ascending one. And the V-I loop will be obvious.

In conclusion, if there is more time for current diffusion of S region to
respond the descending branch than to respond the ascending one, hysteresis
loop in V-I curve, i.e. the HE will be obvious. On the other hand, if the
current diffusion time of S region is similar or the same for the ascending
and descending branch, there will be no obvious V-I loop, namely no HE. The
observation of HE can be controlled by adjusting dI/dt ( the observing time
window ) or d$_S$/d$_W$ ( the distance of current diffusion ) or j$_{cS}$/j$%
_{cW}$ ( the average speed of current diffusion ).

\subsubsection{Magnetic field dependent V-I loop}

It is well known that j$_c$ is dependent on the magnetic field B$_a$.
Commonly, j$_c$ decreases with increasing B$_a$. Therefore, to study the
effect of j$_c$ on the V-I loop is equivalent to study the effect of B$_a$
on it. For the inhomogeneous pinning case, j$_{cW}$ is more sensitive to B$%
_a $ than j$_{cS}$, especially when B$_a$ is not very high \cite{19}. Thus,
we simulated the V-I loops in different applied fields B$_a$, namely
different j$_{cW}$ at fixed j$_{cS}$, dI/dt and d$_S$/d$_W$. Shown in Fig.5
are the numerical results. It is clearly seen that the loop shifts towards
smaller current with decreasing j$_{cW}$ while the area of the loop is
insensitive to j$_{cW}$. In other words, the area of the loop is weakly
affected by magnetic field.

As mentioned above, the existence of HE depends on that the current
diffusion of S region has different responding time for the two sweeping
branches of V-I loop. Only changing j$_{cW}$, i.e. B$_a$, would not affect
the responding time of S region. As a result, the area of V-I loop will not
change acutely.

\section{Experimental}

Because our numerical simulation has obtained more results than the
experimental one reported in reference 12, such as the effect of dI/dt on
the area of loops, we measured V-I curves of Ag-Bi2223 samples in various
applied magnetic fields B$_a$ with different dI/dt to confirm the numerical
prediction. It has been pointed out above that the Ag-Bi2223 sample may be
proper to test our model.

The Ag-Bi2223 samples cut from long silver sheathed tapes come from the
National Center for R\&D on Superconductivity of China. All the samples are
the c-axis textured. The c-axis is perpendicular to the wider surfaces of
the samples. Including the outer silver sheath, the average size of the
samples is approximately: $4mm\times 1mm\times 4cm$. Between each two
neighbor points of four (two voltage contacts and two current contacts) is
1cm in length. The transport measurement was carried out for the sample
immersed in liquid nitrogen. The external magnetic field with range 0 
\symbol{126} 200 Gs was applied in the superconducting state. The current
range of the supply source is 0 \symbol{126} 40 A and sweeping rate range
0.02 \symbol{126} 1 A/s. The measurements were made in the case of B$_a$
parallel to the c-axis of the samples. More details of the sample
preparation and the measurements can be found in reference 18.

The V-I loops were measured by ascending current first and then descending
it. Displayed in Fig.6 are the experimental V-I loops with different dI/dt
and fixed B$_a$. The spottiness of the experimental data came from the
limited picking rate of our measurement system. As reported in reference 12,
clockwise loop of V-I curve was observed. It is clearly seen that dI/dt
affected not only the appearance of V-I curve but also the area of the loop.
For example, the area of loop with dI/dt = 1 A/s is larger than the area of
loop with dI/dt = 0.5 A/s, which is in agreement well with the numerical
result.

Shown in Fig.7 are the experimental V-I loops in different applied fields B$%
_a$ at a fixed dI/dt. It is found that the area of the loop is insensitive
to magnetic field B$_a$ whereas the transport critical current density j$_c$
decreases with increasing applied field very sensitively. This experimental
dependence confirms the above numerical results and thus supports strongly
our model. Furthermore, the agreement between the numerical and experimental
V-I shifting by applied field implies that the critical current density j$_c$
of a sample with spatially non-uniform flux pinning strength, such as the
silver sheathed Bi2223 tapes, is governed mainly by the weaker flux pinning
region (the region with smaller j$_c$). Hence, for these HTS materials,
which have potential for technical applications, the point to enhance their
critical current density is to improve the flux pinning strength of weak
pinning region.

We are unable to measure more V-I curves to test all the numerical curves
for the limited ability of our instrument. For example, dI/dt cannot be too
small because Ohm heat at the current contacts will destroy the sample.

\section{Summary}

We have proposed a model to account for a new kind of history effect in an
inhomogeneous flux pinning superconductor at high temperatures. The model
consists of alternating weak and strong flux pinning regions whose strength
was depicted by different critical current densities j$_{cW}$ and j$_{cS}$,
respectively. Based on the model, our numerical simulation successfully
observed a hysteresis loop in the characteristic V-I curve as reported in
references. Our numerical results also predicted some new characteristics of
the V-I loop at different sweeping rates of applied current dI/dt, applied
magnetic fields B$_a$, the ratio of the two regions' pinning strength and
their thickness.

Physically, due to the difference of flux pinning strength in S and W
regions, the average speeds of flux (or current) diffusion in the two
regions are different too. If dI/dt (or d$_S$/d$_W$, j$_{cS}$/j$_{cW}$) is
proper, the V-I loop, i.e. the HE is obvious. Otherwise, no obvious V-I loop
can be observed.

To confirm the numerical results, electric transport measurement was
conducted to get the characteristic V-I curve of Ag-sheathed Bi$_{2-x}$Pb$_x$%
Sr$_2$Ca$_2$Cu$_3$O$_y$ tapes immersed in liquid nitrogen with or without
applied fields. Clockwise V-I loop was observed to be dependent on dI/dt and
B$_a$. Hence, we presented a new possible mechanism accounting for the new
kind of HE in an inhomogeneous flux pinning superconductor.

\noindent

{\bf Acknowledgment}\ 

The work is supported by the Ministry of Science and Technology of China
(NKBRSF-G1999-0646) and NNSFC (No.19994016).

\noindent
{\large {\bf Figure Captions:}}

\vspace{0.6cm} \noindent
Fig.1. A schematic sketch of the one-dimensional model: a superconducting
slab consisting of periodic strong (S) and weak (W) pinning regions. Vortex
lines enter the system firstly from the surfaces into W region, and then
diffuse from the W into S regions. So the W region surrounds the S region.

\vspace{0.6cm} \noindent
Fig.2.Numerical results, showing the effect of j$_{cS}$/j$_{cW}$ ratio on
the V-I loop (a) and the distribution of current during the ascending (b)
and descending branch (c) of the loop. The arrows indicate the increasing
and decreasing of current, i.e. the time evolution. The dot lines represent
the boundary between S and W regions. d$_S$/d$_W$=1/2, dI/dt=1A/s, j$_{cW}$=2%
$\times $ 10$^8$A/m$^2$.

(2.1) j$_{cS}$/j$_{cW}$ =10, high inhomogeneity of flux pinning. Average
speed of current diffusion in S region is too small comparing with that of W
region, resulting that in S region there is no enough time for current
diffusion in response to both the ascending sweeping and the descending one
at the given parameters. No obvious V-I loop.

(2.2) j$_{cS}$/j$_{cW}$ =4, medium inhomogeneity of flux pinning and lager
average speed of current diffusion in S region, resulting that more current
has penetrated into the S region during the descending branch than the
ascending one at the given parameters. Obvious V-I loop is observed.

(2.3) j$_{cS}$/j$_{cW}$ =1, uniform flux pinning, equal average speed of
current diffusion in the two regions. There is little difference between the
distributions of current during the two branches. No obvious V-I loop.

\noindent
Fig.3. Numerical results, showing the effect of d$_S$/d$_W$ ratio on the V-I
loop (a) and the distribution of current during the ascending (b) and
descending branch (c) of the loop. j$_{cS}$ = 4 j$_{cW}$, dI/dt=1A/s, j$%
_{cW} $=2$\times $10$^8$A/m$^2$.

(3.1) d$_S$/d$_W$ =0.857, the thickness of S region is proper, resulting
that in S region there is enough time for current diffusion in response to
current descending, but the current diffusion is too busy to respond to
current ascending for the given parameters, and area of the V-I loop is
large.

(3.2) d$_S$/d$_W$ =0.222, S region is thinner and the distance of current
diffusion in S region is shorter, resulting that in S region there is much
more time for current diffusion in response to current ascending at the
given parameters, and the area of V-I loop becomes smaller.

(3.3) d$_S$/d$_W$ =0.105, S region is too thin and the distance of current
diffusion in S region is too short, resulting that in S region there is
enough time for current diffusion in response to both the current ascending
and descending at the given parameters, no obvious V-I loop.

\vspace{0.6cm}\noindent
Fig.4. Numerical results. Current sweeping rate dependent V-I loop (a). The
corresponding distributions of current during the ascending (b) and
descending branch (c) were also shown. d$_S$/d$_W$=1/2, j$_{cS}$/j$_{cW}$
=4, j$_{cW}$=2$\times $10$^8$A/m$^2$.

(4.1) dI/dt=0.01A/s, the current sweeping rate is so slow that in S region
there are enough time for current diffusion in response to both the
ascending and descending current sweeping at the given parameters, no
obvious V-I loop.

(4.2) dI/dt=2A/s, the current sweeping is faster, resulting that in S region
there is no enough time for current diffusion in response to current
ascending at the given parameters, and the V-I loop is larger.

(4.3) dI/dt=10A/s, the current sweeping is so fast that in S region there is
no enough time for current diffusion in response to both the ascending and
descending current sweeping at the given parameters. No obvious V-I loop.

\vspace{0.6cm}\noindent
Fig.5. Numerical results. Magnetic field dependent V-I loop, where the area
of the loop is insensitive to j$_{cW}$ while the transport j$_c$ is strongly
affected by magnetic field. d$_S$/d$_W$=1/2, j$_{cS}$ =8$\times $10$^8$A/m$%
^2 $, dI/dt=1A/s.

\noindent
Fig.6. Experimental V-I loops of the Ag-Bi2223 tape with different current
sweeping rates dI/dt, which is qualitatively with the numerical curves shown
in Fig.4 and is a conformation of our model.

\vspace{0.6cm}\noindent
Fig.7. Experimental V-I loops of the Ag-Bi2223 tape at different applied
magnetic fields, which is qualitatively with the numerical curves shown in
Fig.5. It is noted that the area of the loop is insensitive to B$_a$ whereas
the value of the j$_c$ is strongly affected by magnetic field, which
supports the present model further.

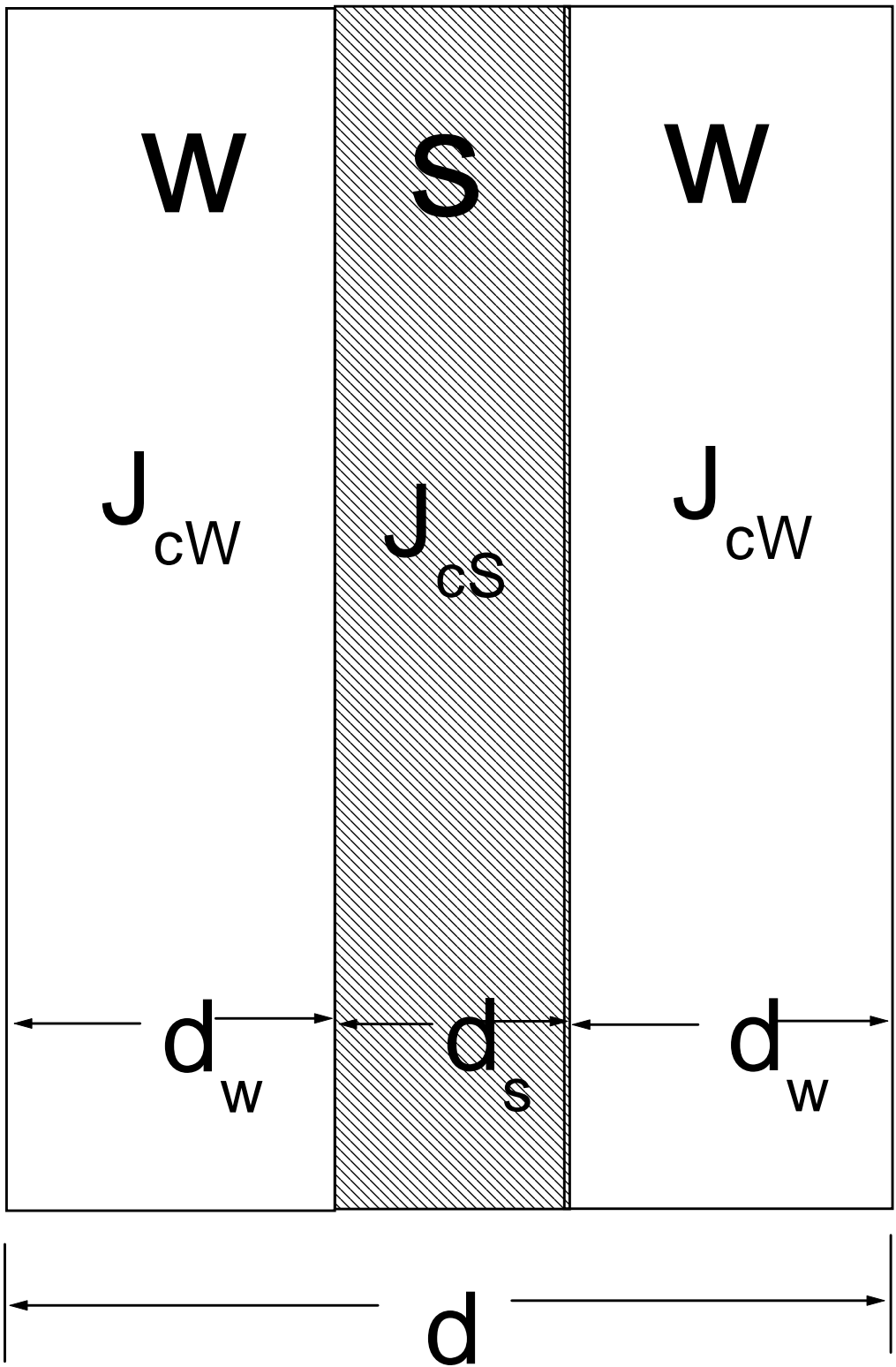
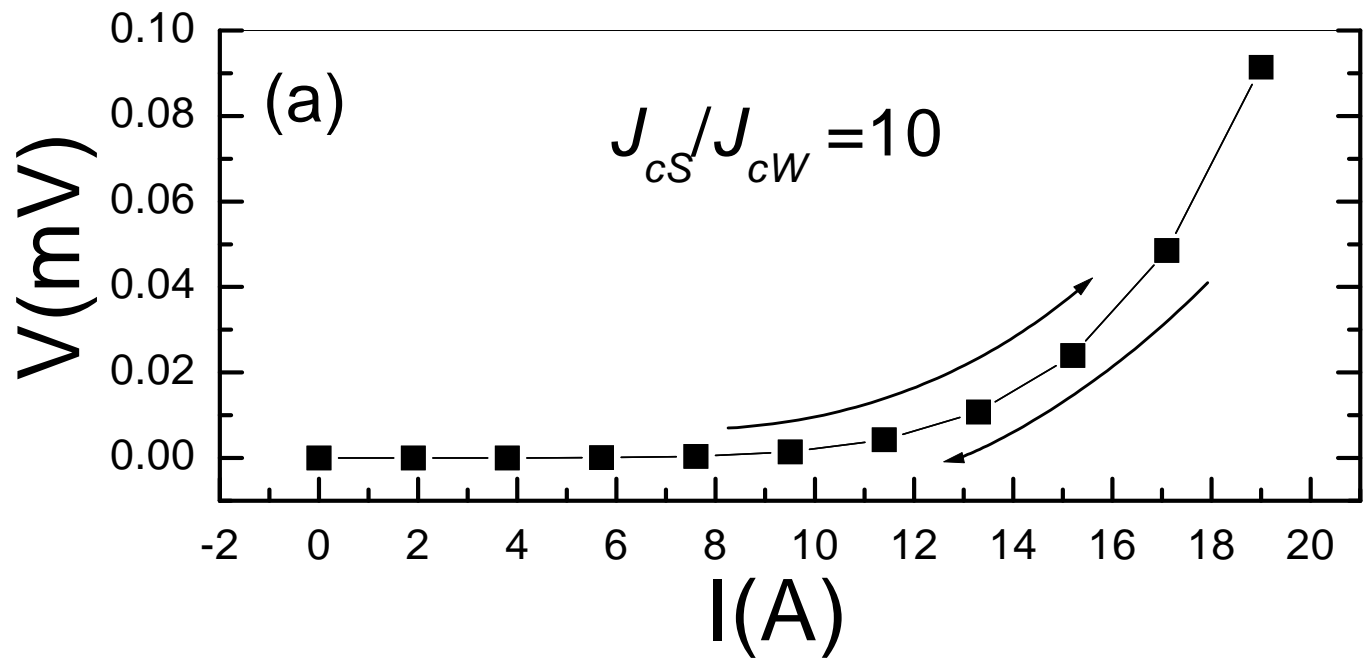
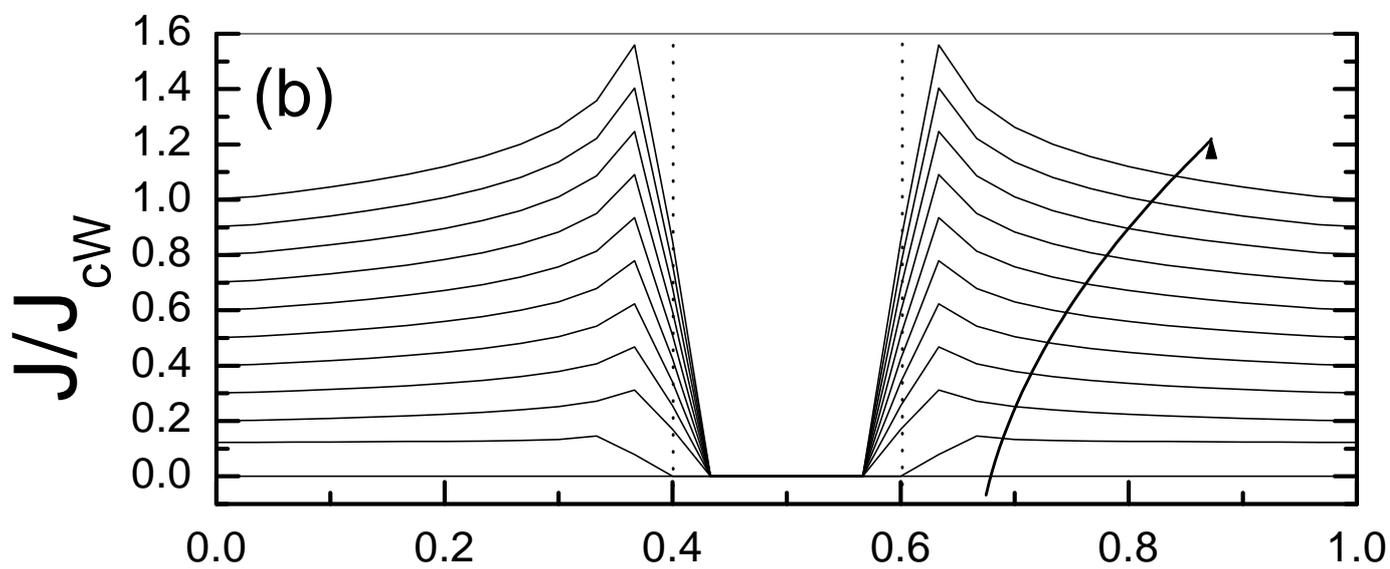
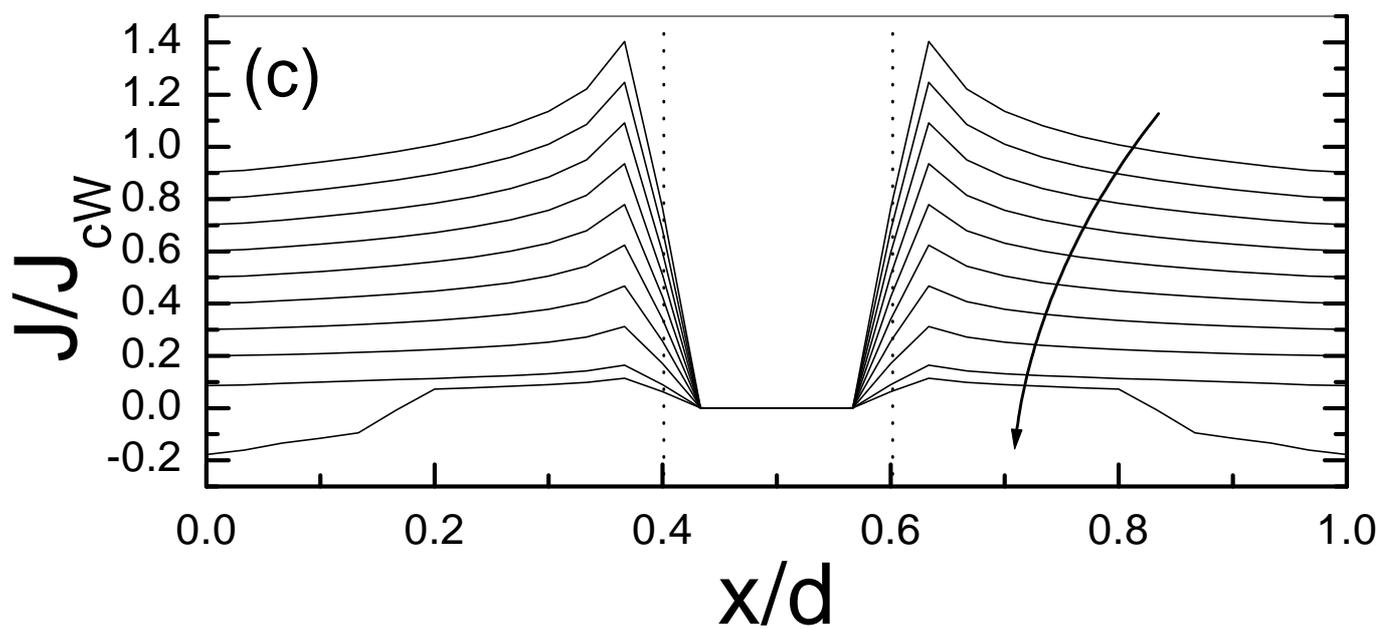
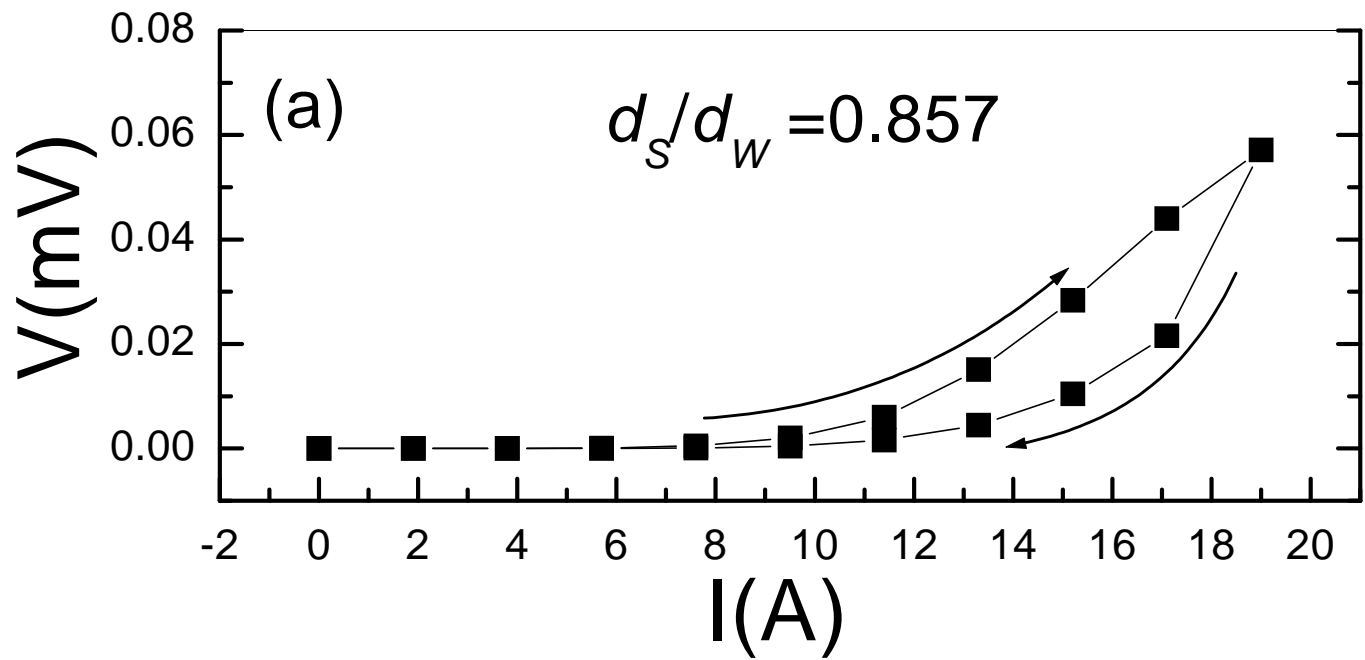
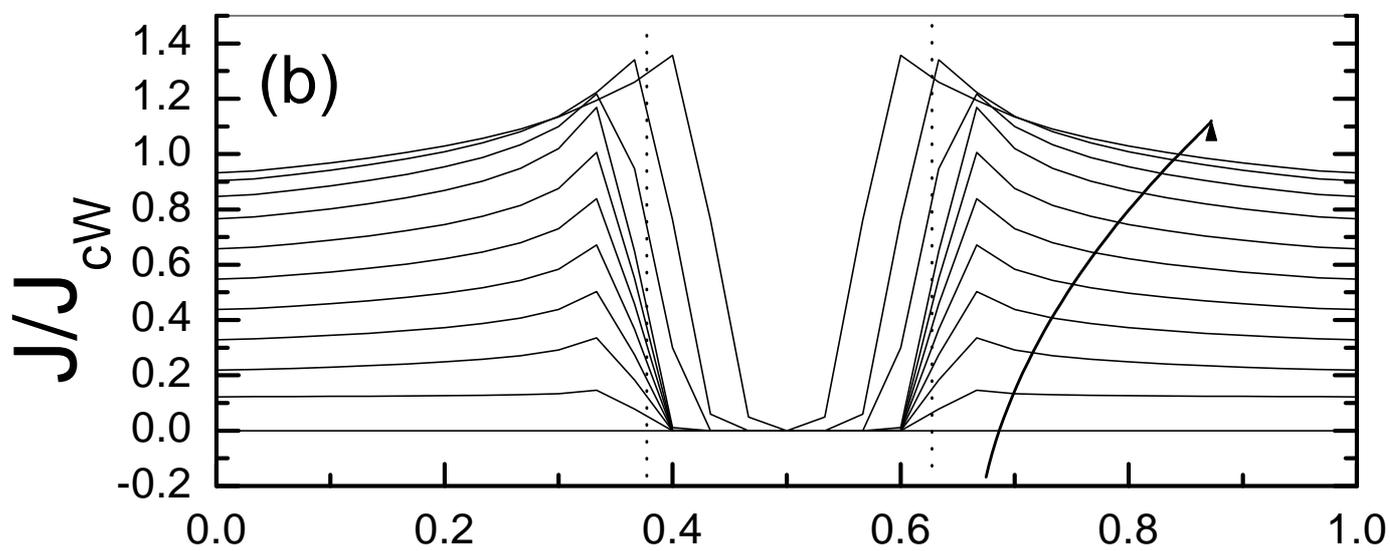
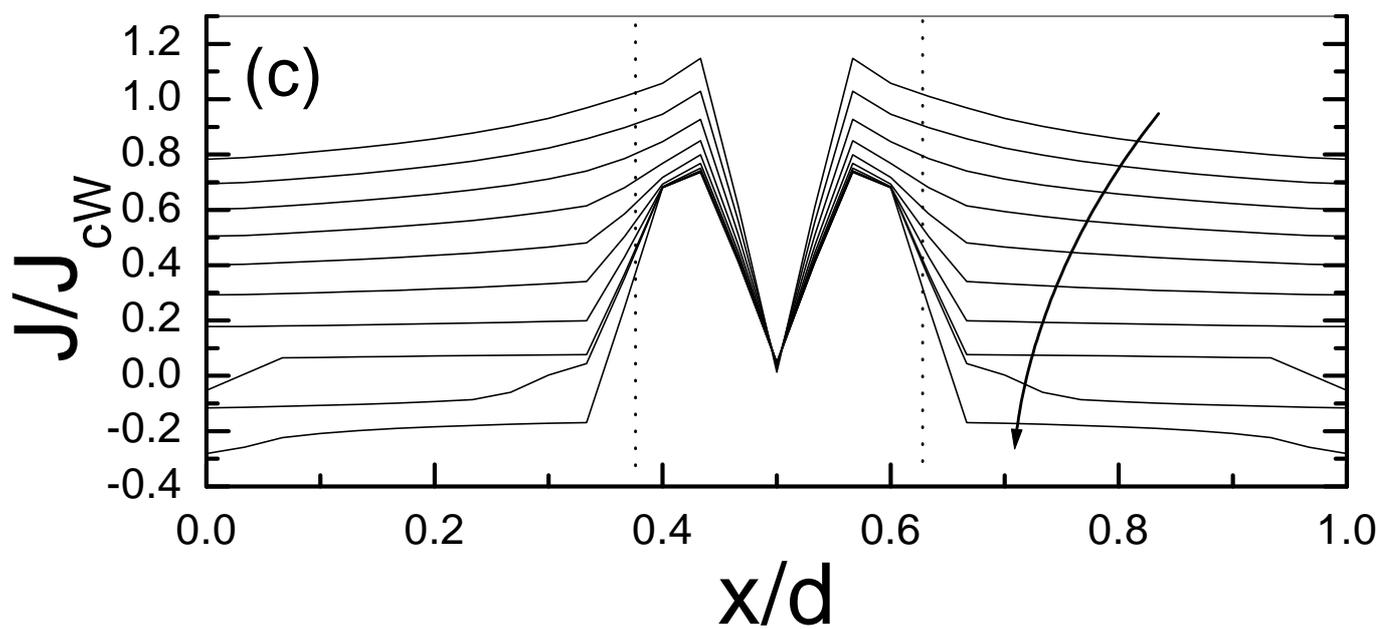
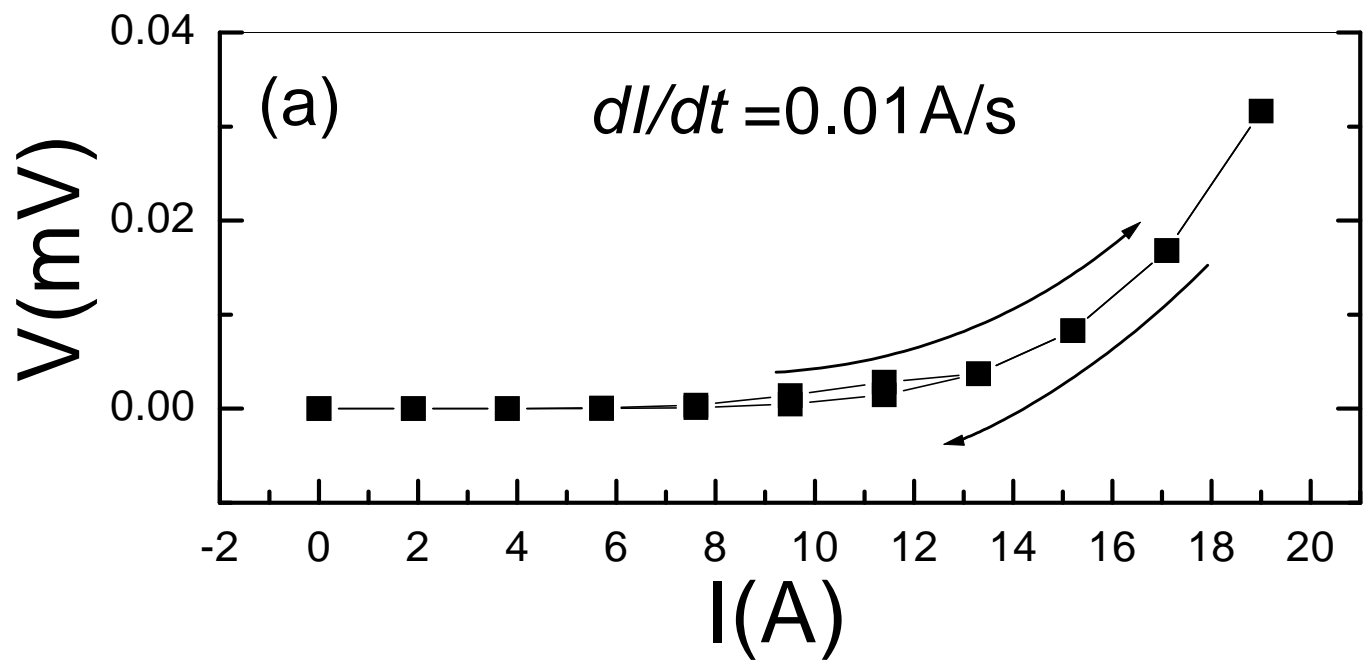
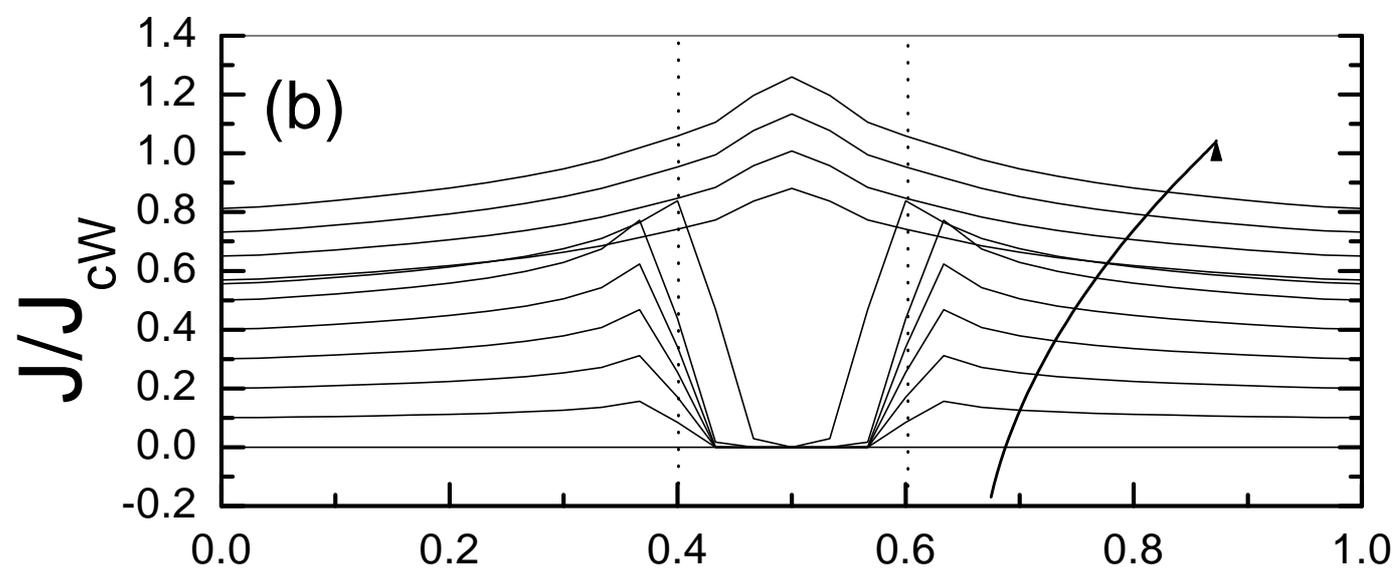
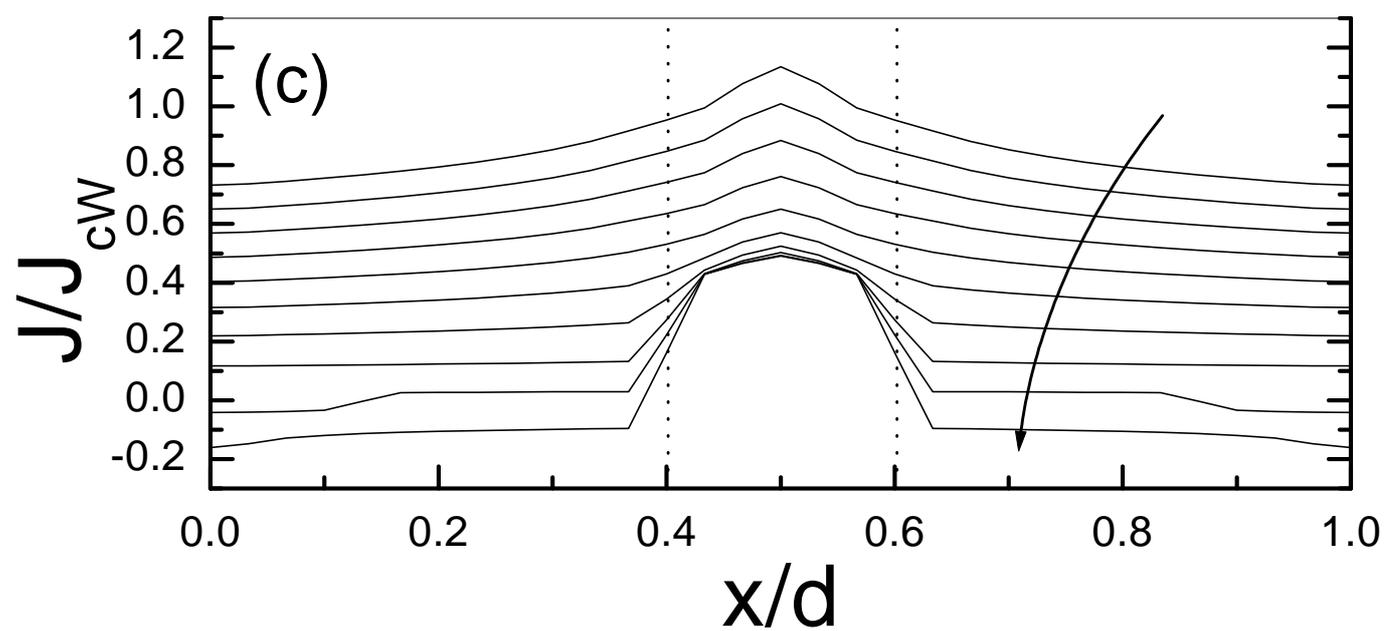
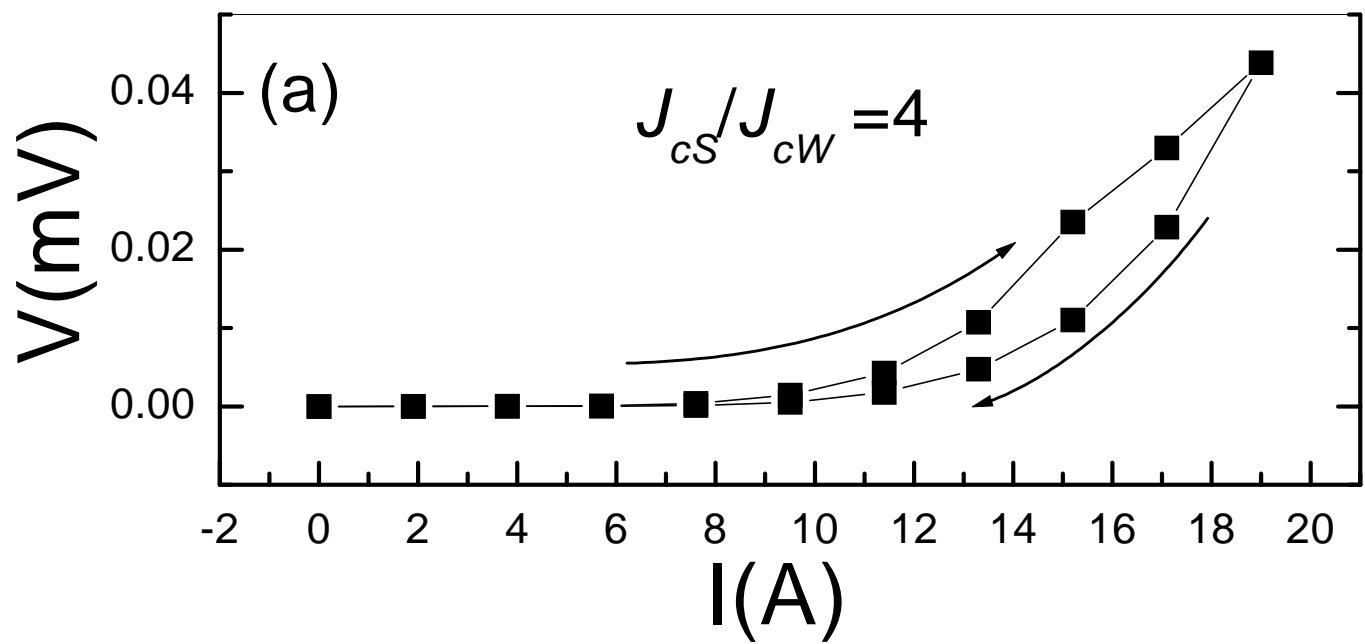
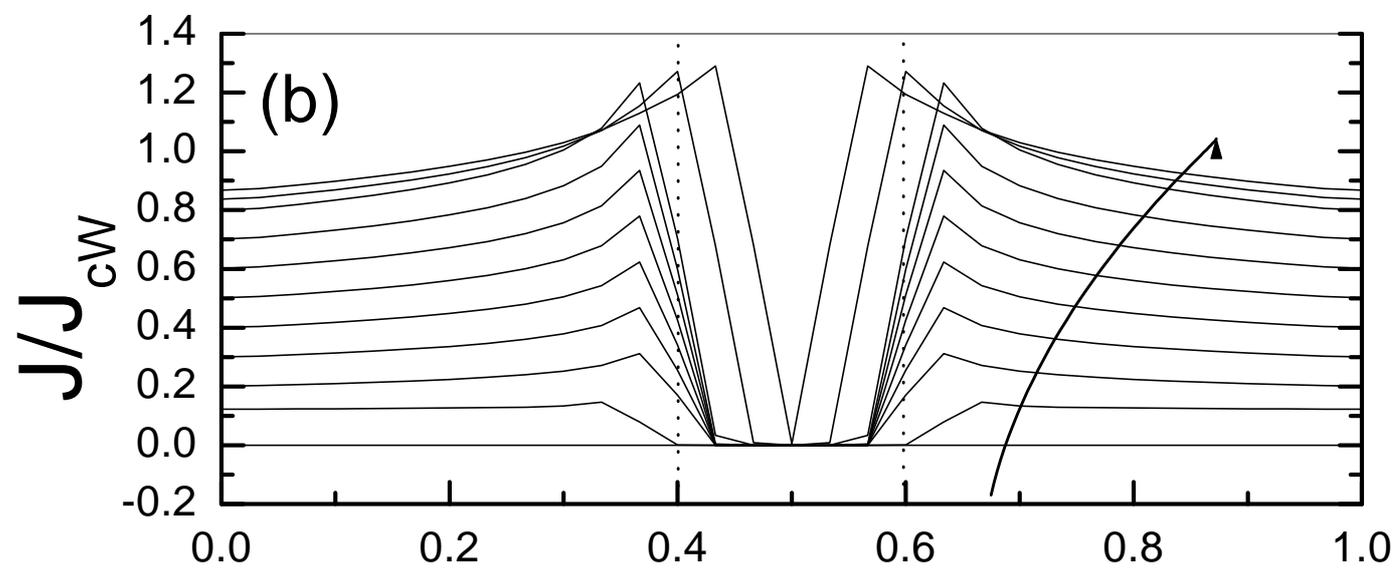
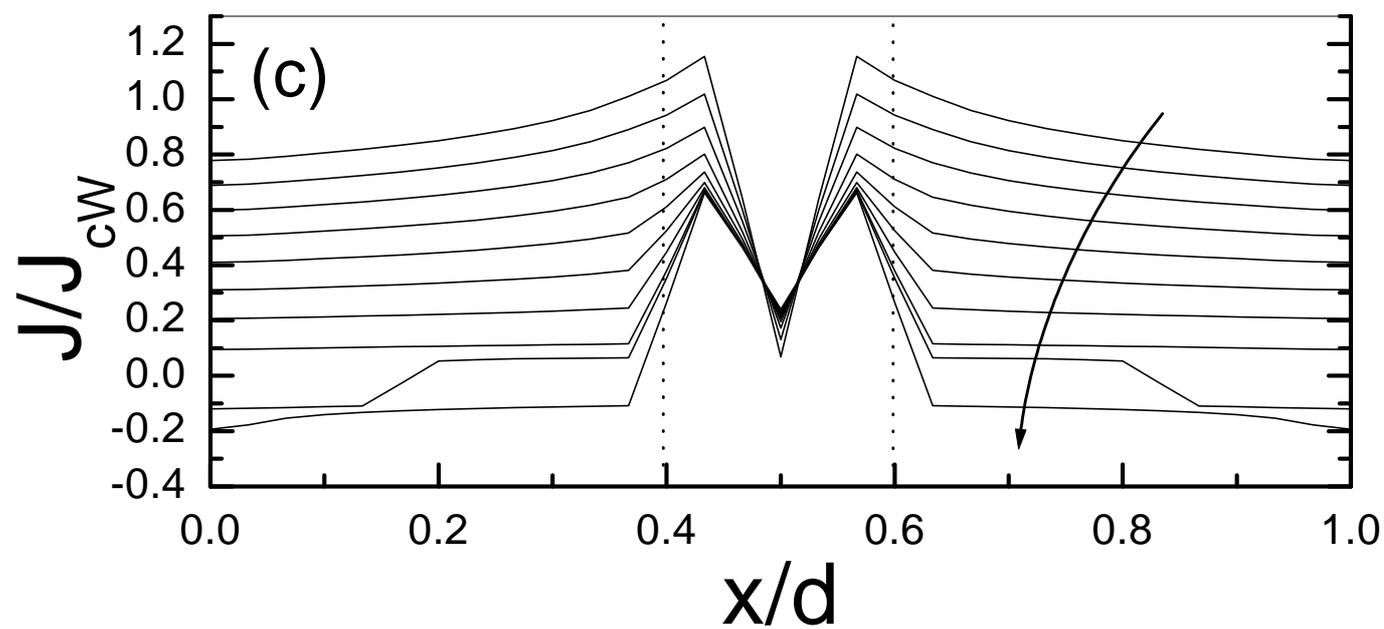
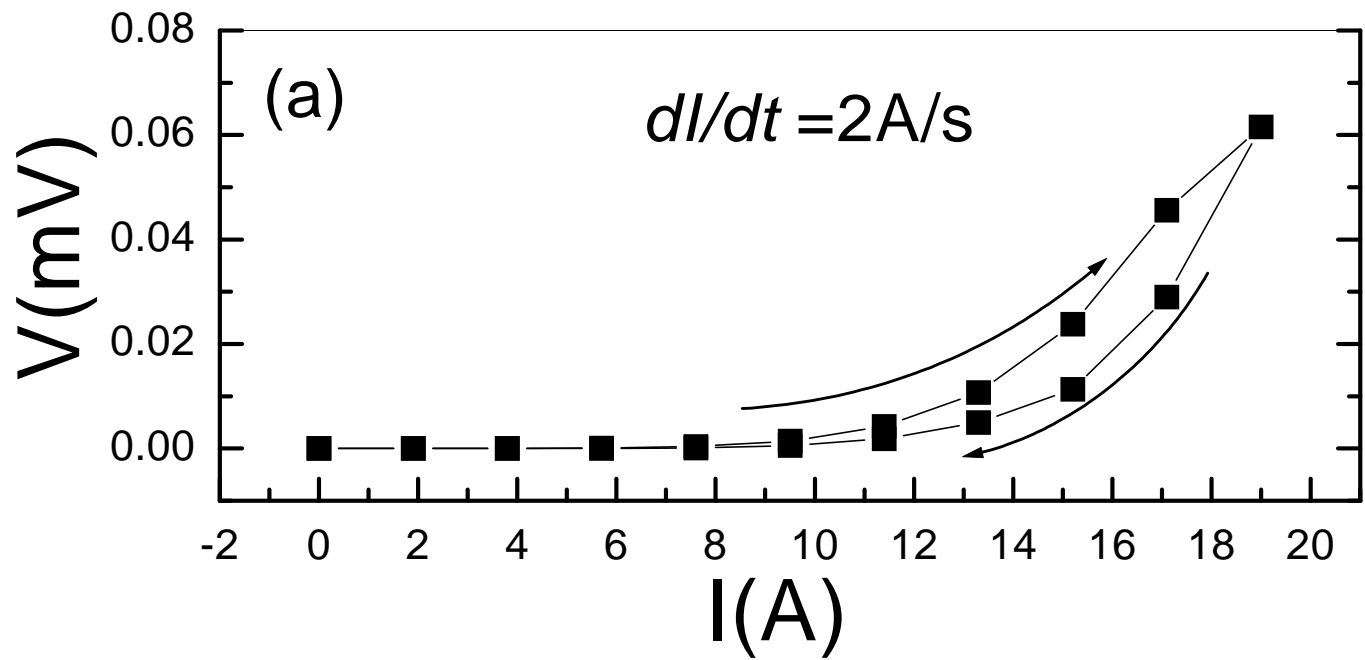
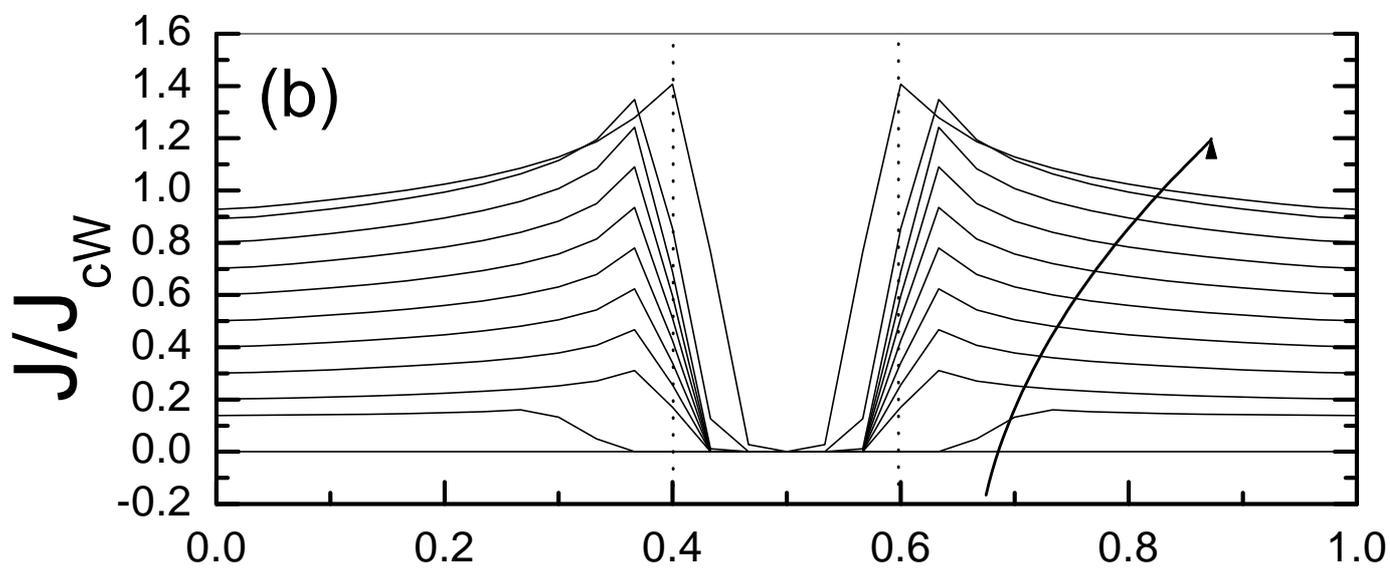
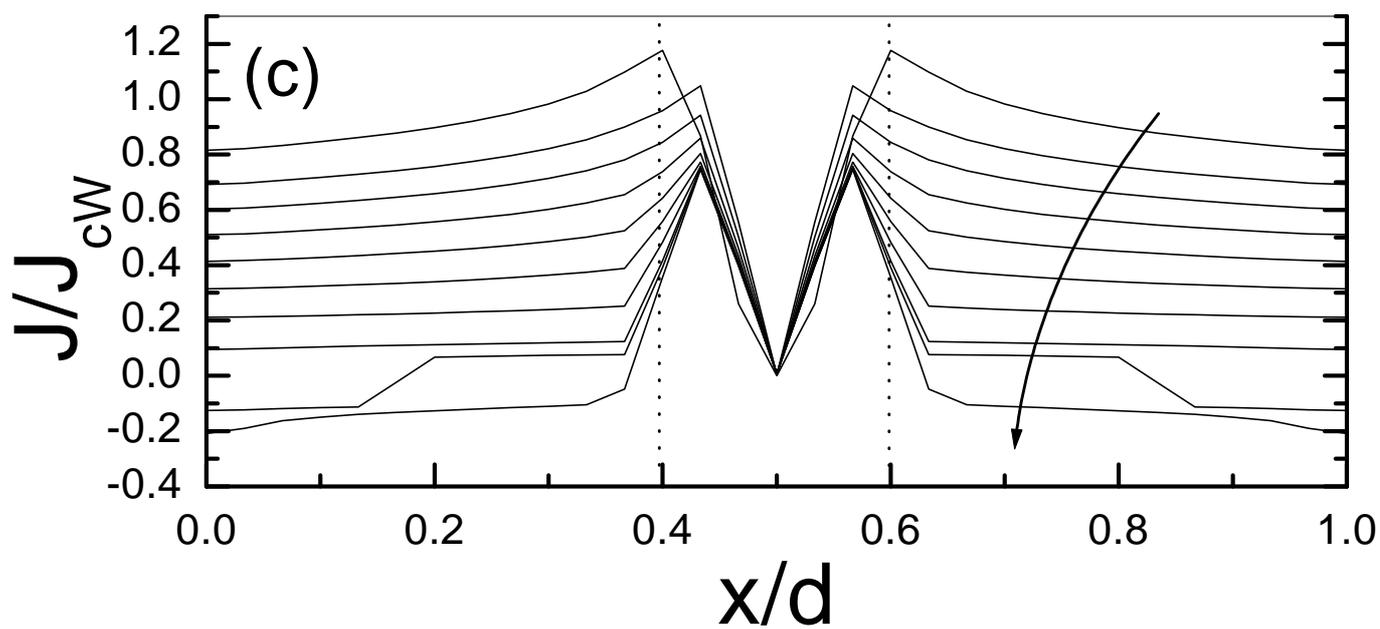
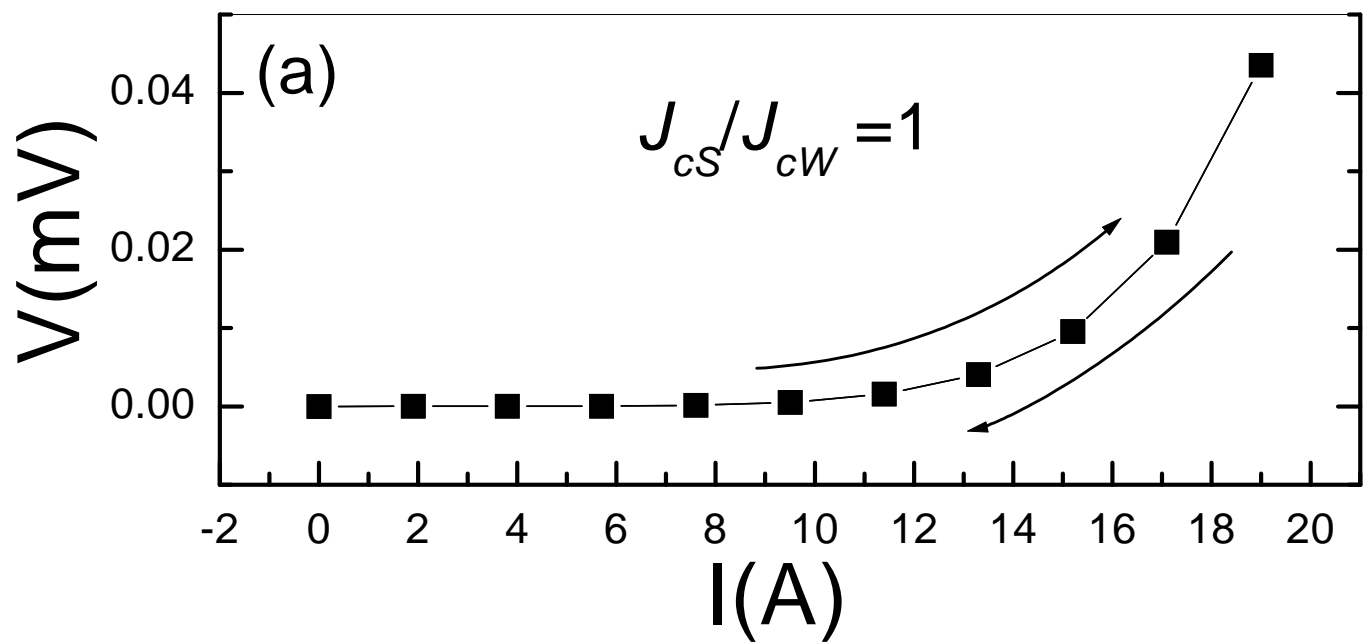
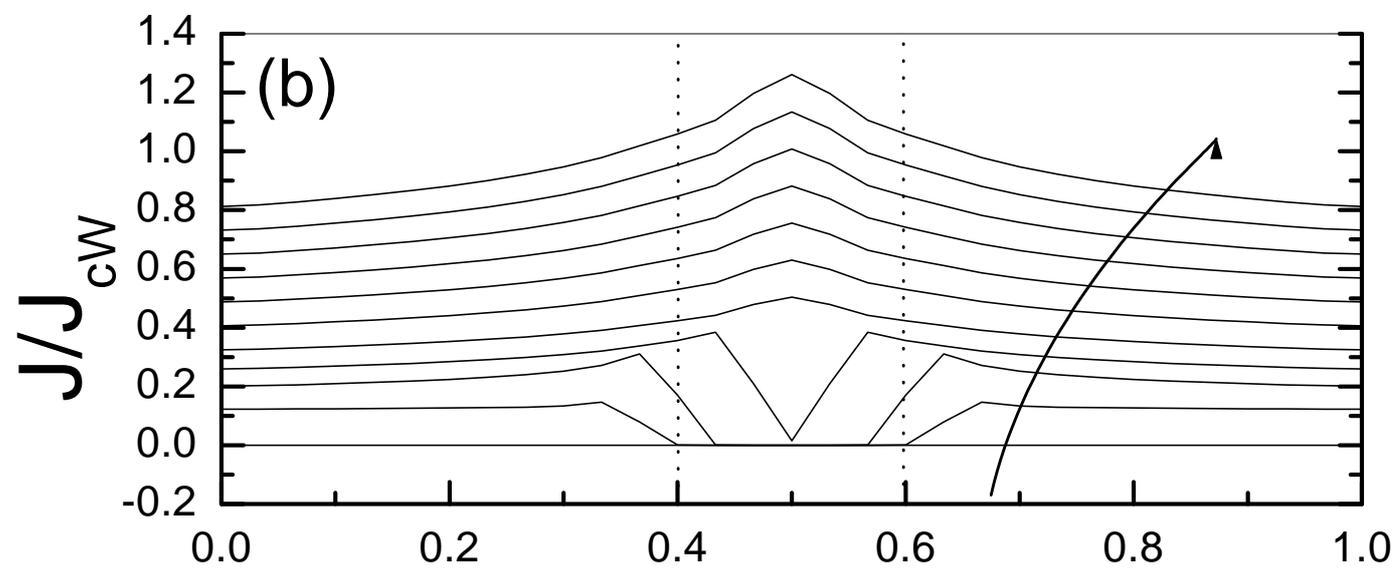
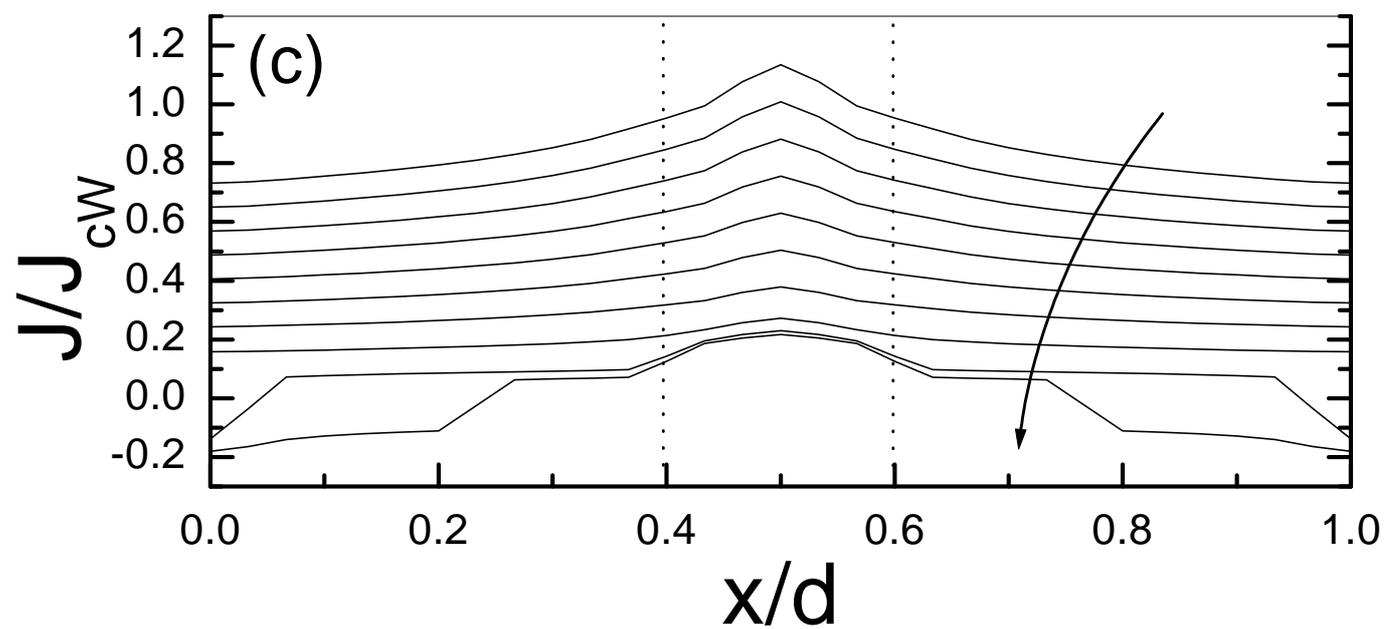
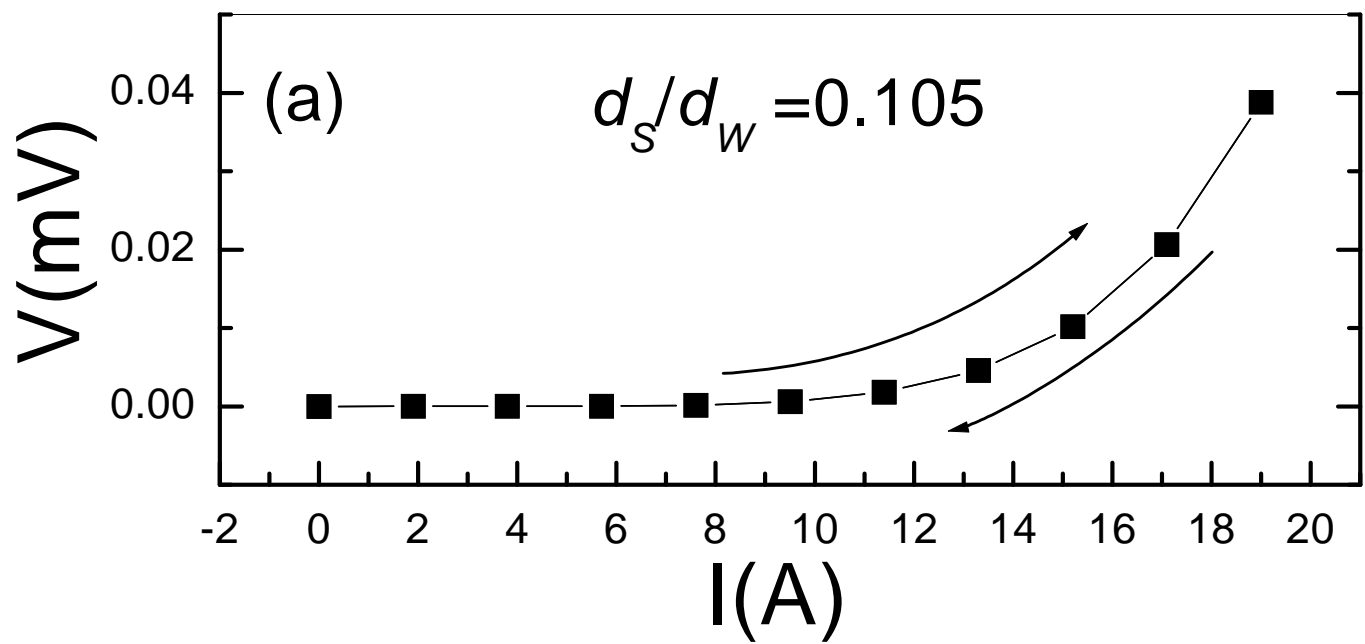
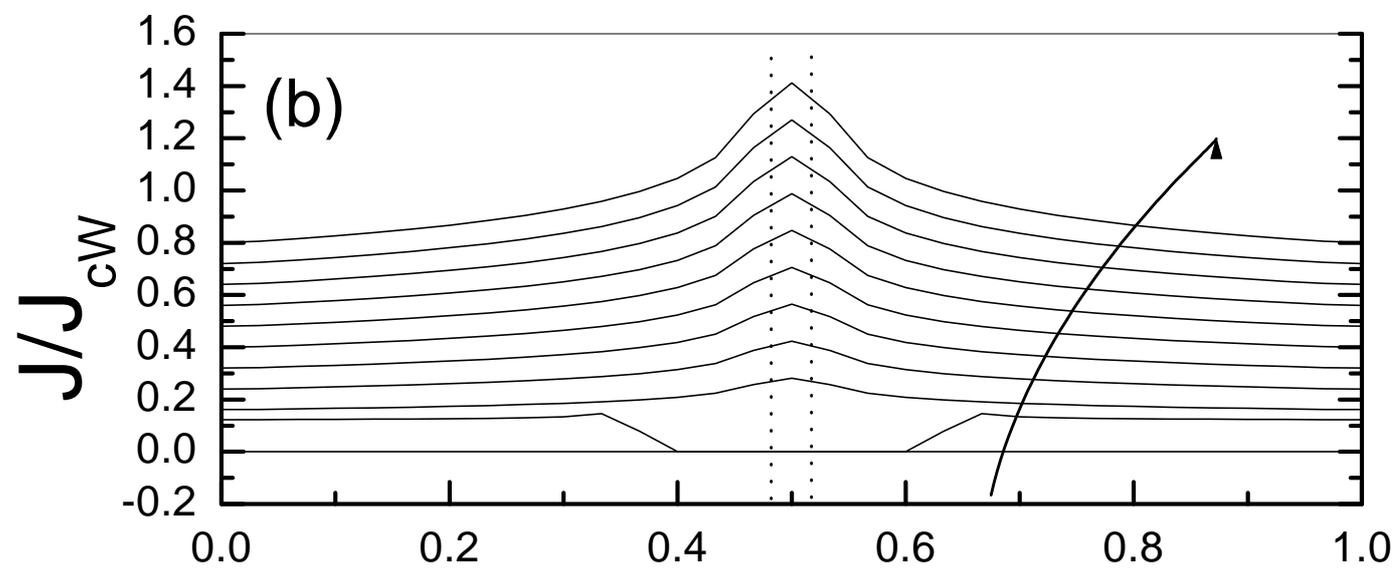
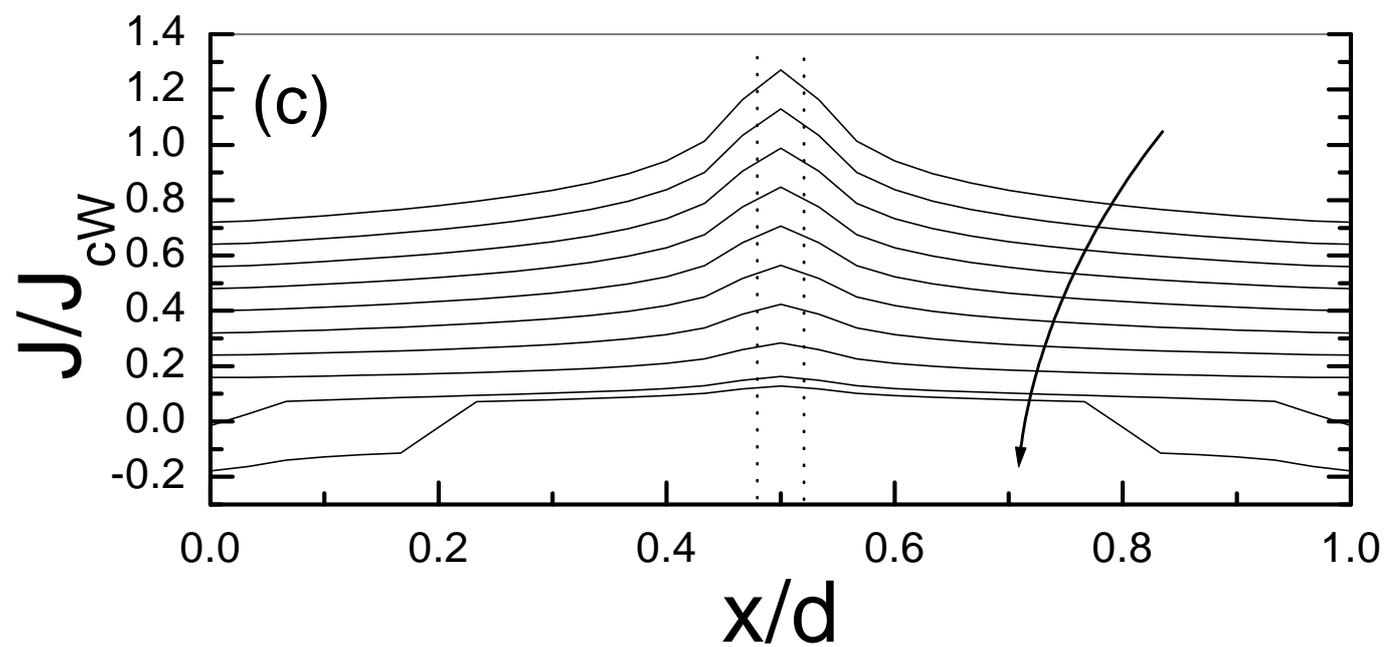
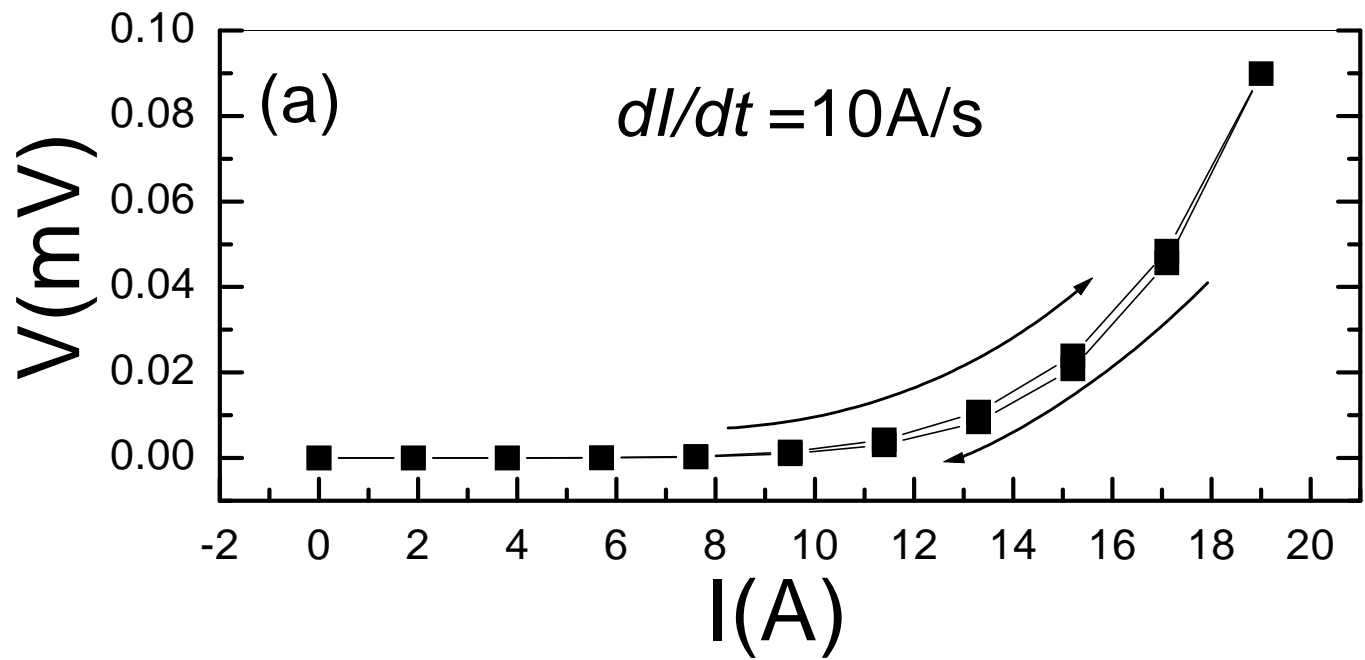
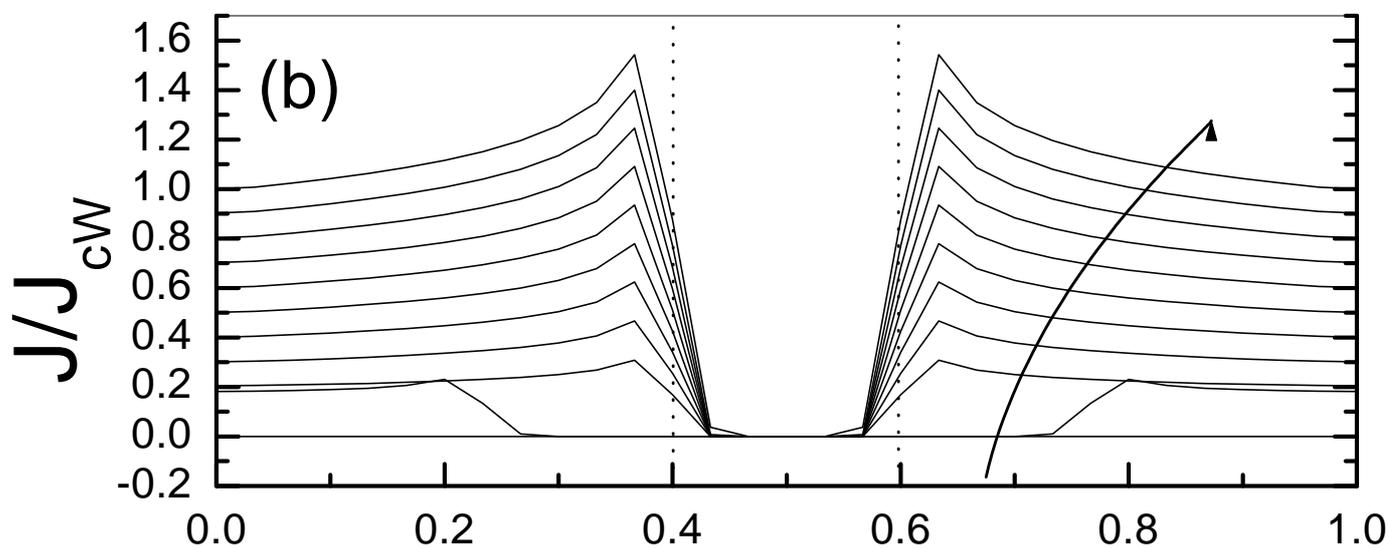
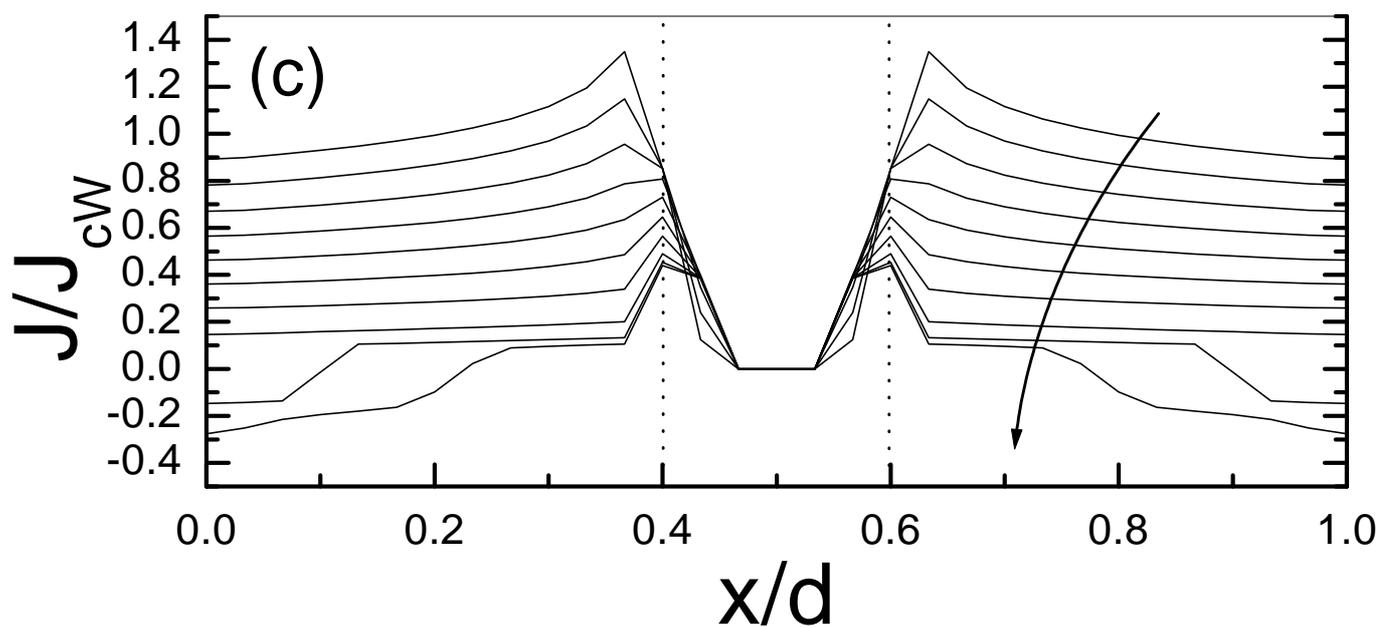
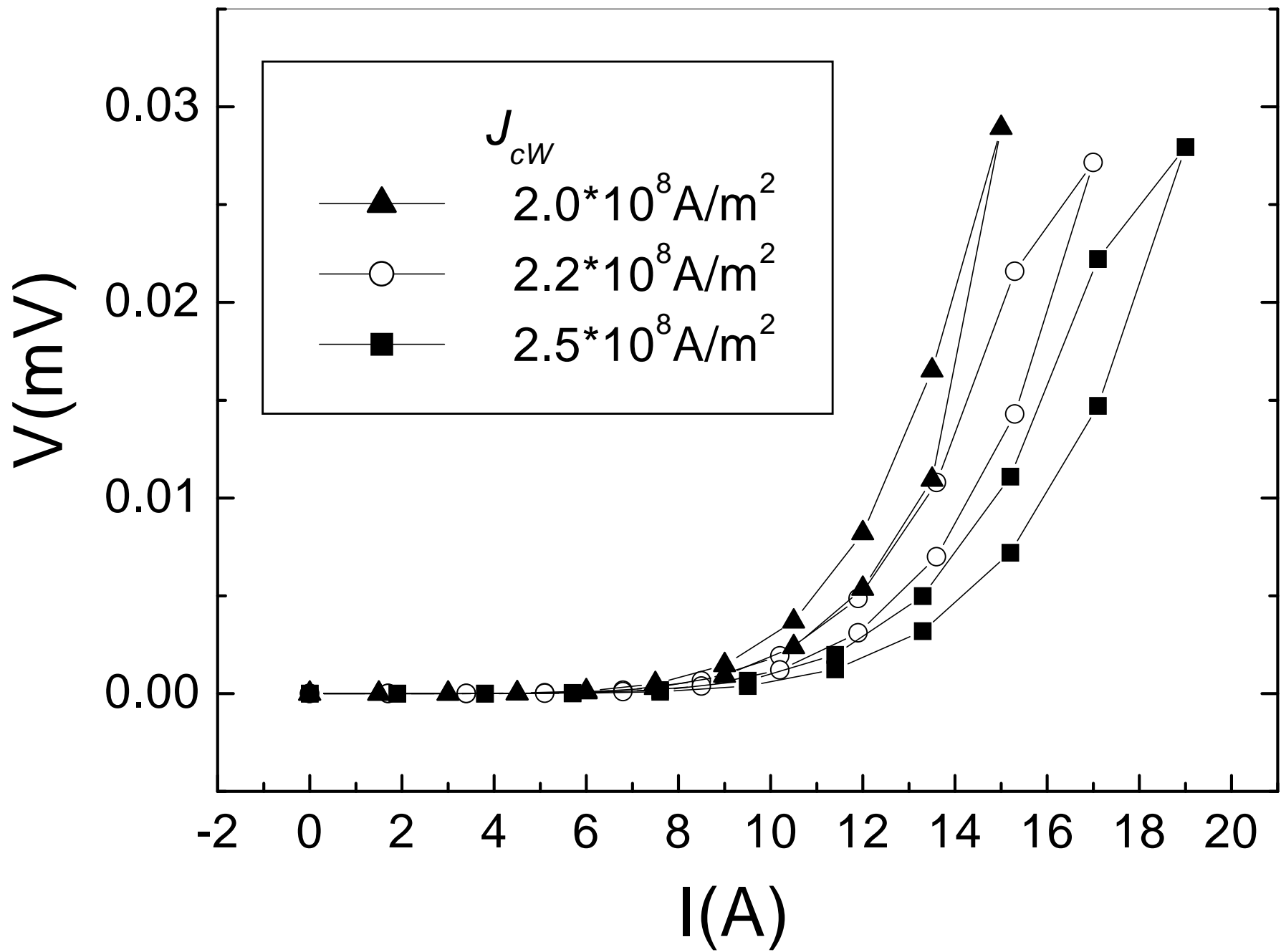
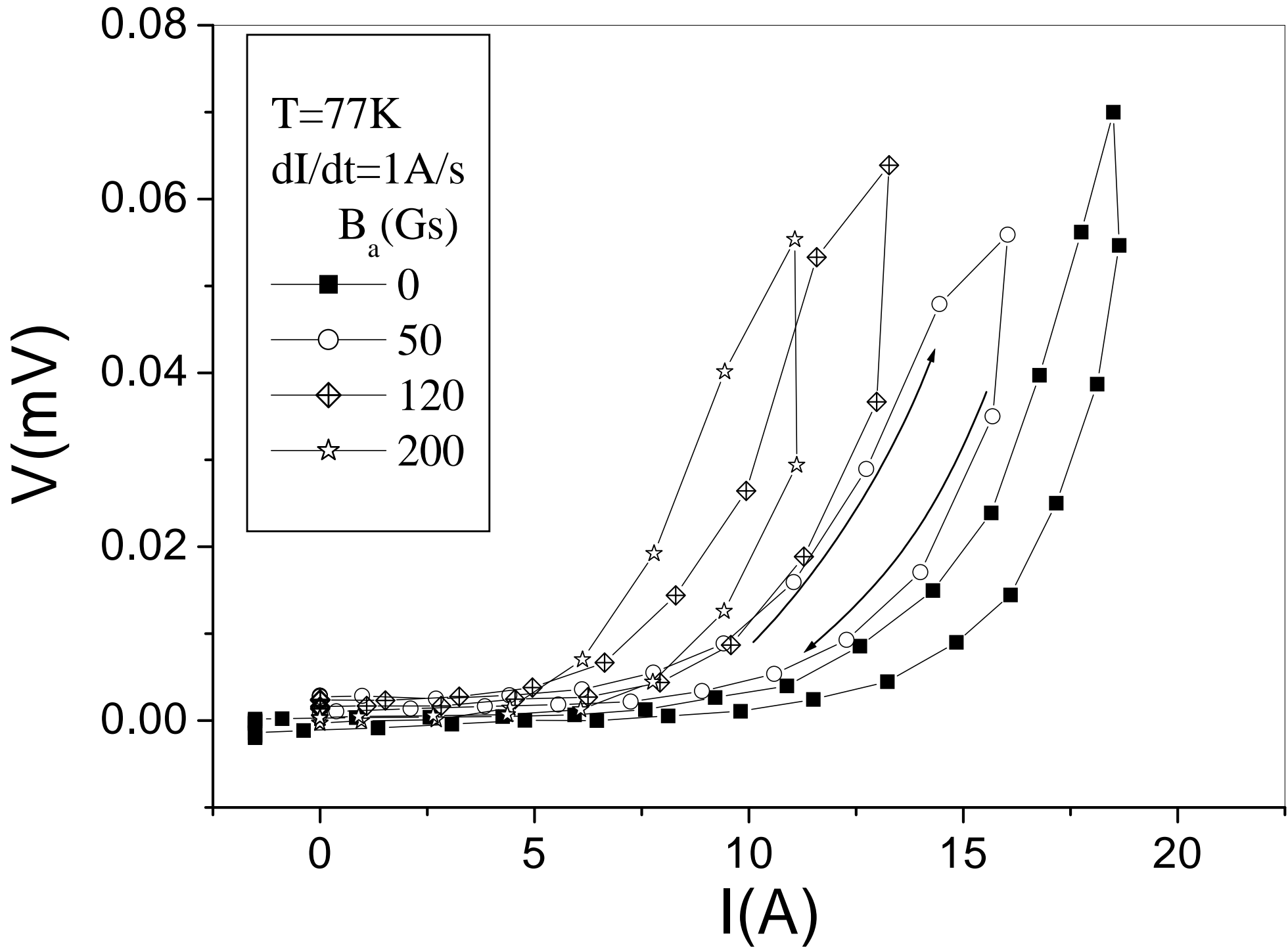
\end{document}